\documentclass[letter]{aa} 

\usepackage{graphicx}
\graphicspath{{./}{figures/}}

\usepackage{txfonts}
\usepackage{hyperref}
\usepackage{color}
\usepackage{multirow}
\usepackage{natbib}
\bibpunct{(}{)}{;}{a}{}{,} 

\usepackage{ulem}

\begin{document} 


\title{Eddington bias for cosmic neutrino sources}

\author{Nora Linn Strotjohann
        \inst{1}
    \and
    Marek Kowalski
    \inst{1}\inst{2}
    \and
        Anna Franckowiak
    \inst{1}
}

\institute{Desy Zeuthen, Platanenallee 6, 15738 Zeuthen, Germany\\
    \email{nora.linn.strotjohann@gmail.com}
    \and
    Institut f\"ur Physik, Humboldt-Universit\"at zu Berlin, 12489 Berlin, Germany
}


\abstract{
We describe a consequence of the Eddington bias which occurs when a single astrophysical neutrino event is used to infer the neutrino flux of the source. A trial factor is introduced by the potentially large number of similar sources that remain undetected; if  this factor is not accounted for the luminosity of the observed source can be overestimated by several orders of magnitude. Based on the resulting unrealistically high neutrino fluxes, associations between high-energy neutrinos and potential counterparts or emission scenarios were rejected in the past. 
Correcting for the bias might justify a reevaluation of these cases.}


\maketitle




\section{Introduction}
Neutrinos interact weakly, and hence are only detected  in small numbers. This is  particularly true for high-energy neutrino astronomy, where for decades it has been established wisdom that ``neutrino physics is the art of learning a great deal by observing nothing'' (Haim Harari; see \citealt{quotes}). With the first detections of high-energy cosmic neutrinos \citep{hese, muondiffflux} this is gradually changing, and yet the available low statistics leaves its imprints. Several associations between individual high-energy events and potential electromagnetic counterparts have been suggested, the most significant being the detection of a single high-energy event in coincidence with a very high-energy gamma-ray flare from the blazar TXS\,$0506+056$ \citep{txsmmpaper}. A subsequent search for additional neutrino emission from this source revealed a flare of $13\pm5$ lower energy events several years earlier \citep{txsneutrinopaper}.
Here, we explore what a single event can and cannot reveal about the neutrino flux of a source by studying a population of simulated sources.

An everyday example of a similar bias could be the attention that a lottery winner receives in a local newspaper. Without knowing the details of the game, a reader might get the false impression that the chances of winning are relatively high. However, to obtain a realistic estimate of the odds, the large number of players who did not win (and were not mentioned in the newspaper) has to be taken into account.
In neutrino astronomy, a single neutrino event detected from a known source corresponds to the lucky winner. It is, however, crucial to consider the potentially large trial factor introduced by similar sources from which no event is detected. Due to this trial factor a population of numerous faint sources may contribute significantly to the astrophysical neutrino flux, even if a detection is unlikely for every individual source (as has been quantified for resolved and unresolved blazars by \citealt{krauss2015} and \citealt{palladino2018}).

The bias described here can be considered the low statistics case of the Eddington bias~\citep{eddington1913}. Eddington described how the rate of rare, bright stars is overestimated when a fraction of objects from a more numerous fainter population is wrongly associated with the brighter class due to the error on the flux measurement. When searching for neutrino sources, a large population of sources have fluxes below the detection threshold of the IceCube neutrino observatory \citep{detectorpaper}; nevertheless, a few of them might pass the threshold due to statistical overfluctuations. This enhances the number of sources observed just at the detection threshold as also seen for gamma-ray sources by the \textit{Fermi}-LAT telescope \citep{ackermann2016}. Another consequence of the Eddington bias is that the flux of neutrino source candidates is systematically overestimated. The size and implications of this bias on the neutrino flux are explored in the following.

\section{Quantifying the bias}
\label{sec:bias}

Contrary to the lots in a lottery, neutrino sources within a population do not all have the same probability of being detected. Instead, they follow a distribution depending on the density of sources, their evolution and luminosity function, and on the direction dependent acceptance of the detector. To account for the different source fluxes, we simulate a population of equally luminous neutrino point sources that are evenly distributed throughout the universe out to a redshift of $z=4$. Their flux on Earth is calculated for an $E^{-2}$ spectrum and using the cosmological parameters from \citet{planck2016}.

\begin{figure*}[tbp]
\begin{center}
\includegraphics[width=0.9\textwidth]{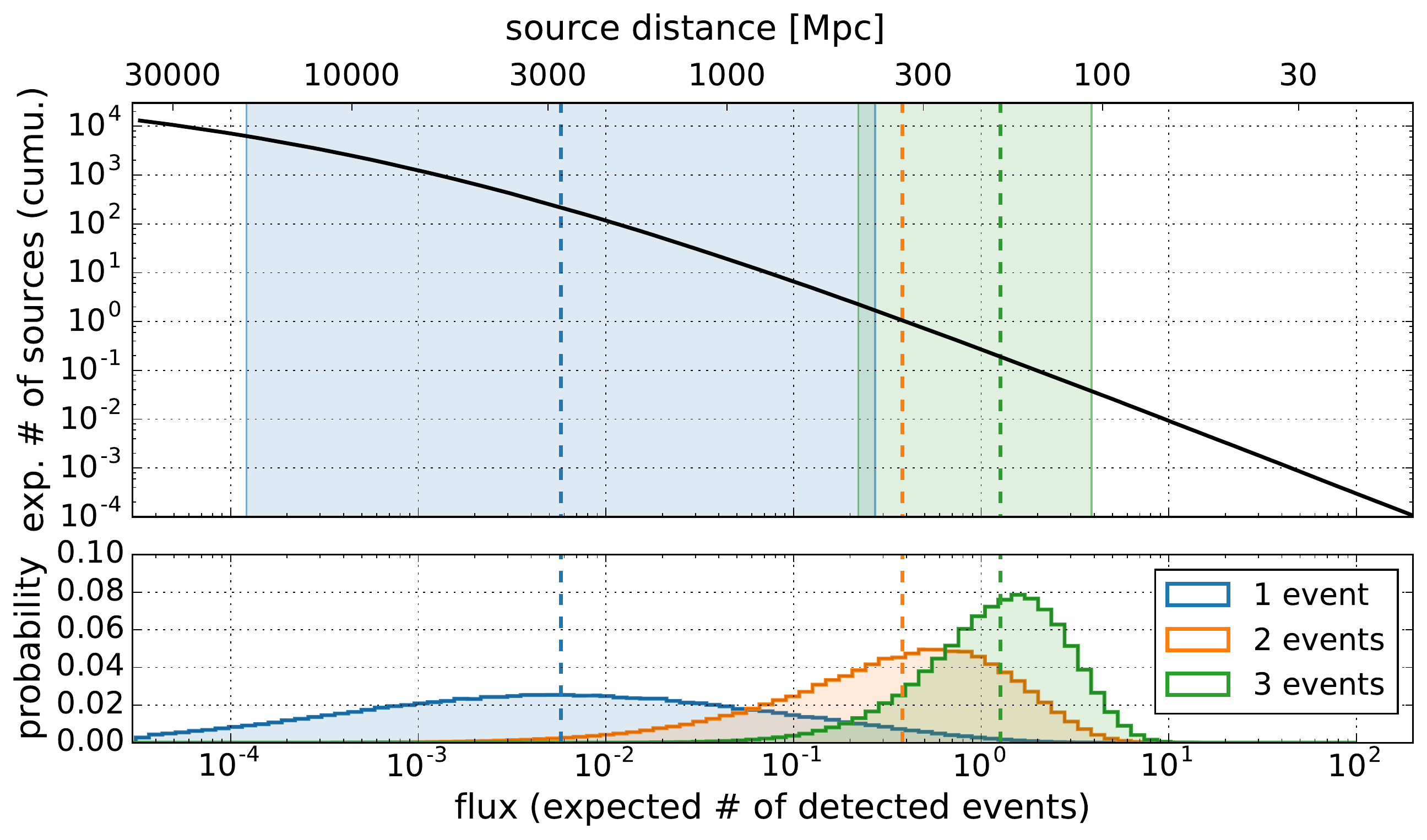}
\end{center}
\caption{LogN-LogS distribution of the simulated sources. A constant source density of $8\times10^{-9}\,\text{Mpc}^{-3}$ (the effective density of BL Lac objects; see Sect.~\ref{sec:zdistributions}) corresponds to $1.2\times10^4$ sources within redshift $z=4$. The flux on the x-axis is given as the expected number of detected neutrino events and is normalized such that ten events are expected from the complete population. Since all sources are equally luminous, the flux can be converted to the source distance shown on the upper x-axis. For nearby sources, the LogN-LogS distribution is proportional to $S^{-3/2}$ as expected for an Euclidean universe. The probability distributions in the lower panel show from which sources the detection of one, two, or three events is most likely expected.
The dashed lines indicate the median source flux and the colored bands in the upper panel include 90\% of the probability distributions for one and three detected events. In the adopted example, a source detected with a single event is most likely located at a distance between 0.5 and 20\,Gpc and its flux can be as small as $10^{-4}$ expected events. \label{fig:flux_dist}}
\end{figure*}

The black line in the upper panel of Fig.~\ref{fig:flux_dist} shows the LogN-LogS distribution of the simulated source population, i.e.,  the flux per source versus the cumulative number of sources. In this example the number of sources in the universe is set to $1.2\times10^4$ within $z=4$, which corresponds to a density of $8\times10^{-9}\,\text{Mpc}^{-3}$, the effective blazar density. Cosmic source evolution is neglected in this example, but the use of effective source densities allows us to approximately estimate the size of the bias as verified in Sect.~\ref{sec:zdistributions}. The steeply falling black line in Fig.~\ref{fig:flux_dist} illustrates that the geometry of the universe yields a large number of faint sources and few bright sources.

The flux per source is given as the expected number of detected events. For simplicity, we assume a generic neutrino detector that is equally sensitive in all directions. In this study we quantify how likely it is that one or several astrophysical neutrino events are detected from the different sources within the simulated population. We do not consider other properties of the neutrino detector or a specific analysis, such as the detected number of background events or the angular resolution.
The flux of the complete source population  in this example is normalized to ten neutrino events, which approximately corresponds to the number of astrophysical extremely high-energy (EHE) events expected within three years of IceCube data \citep{realtimepaper}. The neutrino emission from blazars has been restricted to $<\!27$\% of the detected flux \citep{blazarlimit}, so ten or fewer EHE events are expected from blazars within ten years. 

For each simulated source, the Poisson probability of observing one, two, or three events is calculated;  the resulting probability distributions are shown in the lower panel of Fig.~\ref{fig:flux_dist}. For each source $i$, we hence calculate the term $d P_i(\mu_i, k)/d \mu_i$, where $\mu_i$ is the expected flux on Earth; $k=1,2,3$ is the number of detected events; and $P_i(\mu_i, k)$ is the Poisson probability of detecting $k$ events. The contributions from all sources are added and the resulting distributions are normalized to one. They thus show from which sources in the population the detected neutrino signal most likely originates. While brighter sources have a higher individual probability of being detectable, they are rare, and the much larger number of fainter sources might be more likely to yield a detection. The dashed lines show the median flux of a source detected with one, two, or three events and the shaded bands in the upper panel of Fig.~\ref{fig:flux_dist} contain 90\% of the probability distributions.

The median flux of a source detected with a single event (shown as the blue dashed line in Fig.~\ref{fig:flux_dist}) is much smaller than one expected event. For the assumptions used here the median flux is close to $0.006$, which corresponds to the 220th brightest source in the simulated population. For this example there is only a $0.8\%$ chance that a source detected with one event has an expectation value of one or higher, and hence is ruled out at $\sim99\%$ confidence level. It is thus unlikely to detect a single event from one of the brightest sources; instead, the many fainter sources have a higher probability of producing such a signal.

\section{Impact of cosmic source evolution}
\label{sec:zdistributions}

\begin{table*}
{\small
\hfill{}
\caption{Size of the bias for different redshift distributions and luminosity functions. For each source class the reference to the used redshift distribution, the source density at $z=0$, and the resulting number of sources is given. In addition, two different luminosity functions were tested. The first line represents the standard candle scenario, while in the second line the calculation was repeated assuming rather large luminosity variations described by a lognormal distribution with a width of one order of magnitude. The 90\% confidence region for one detected event is characterized by quoting the median and the 5\% and 95\% percentiles of the probability distribution. To first order the effect of the redshift distributions and luminosity variations can be absorbed into an effective source density. For these densities the bias is equally large (same median) for a population of standard candle sources without evolution (see Fig.~\ref{fig:comparison}). The effective density for BL Lac objects without luminosity fluctuations, $8\times10^{-9}\,\text{Mpc}^{-3}$, is used for the calculation in Sect.~\ref{sec:bias}. As before fluxes are given as the expected number of detected events and the emission of the total population is normalized to ten detected events.}
\label{tab:effective_densities}     
\begin{center}
\begin{tabular}{lccccccc}     
\hline\hline       
{\bf source class} &
{\bf source density} &
{\bf \# sources} &
{\bf lumi. variations}&
\multicolumn{3}{c}{{\bf flux of source det. with one event}} &
{\bf eff. density} \\ 
       &
(at $z=0$) &
(within $z<4$) &
width of gaussian &
5\% perc. &
median &
95\% perc. &
\\ 
       &
$[\text{Mpc}^{-3}]$ &
&
[dex]&
&
&
&
$[\text{Mpc}^{-3}]$ \\ 
\hline
\noalign{\vskip 2mm}
{\bf FSRQs} &  \multirow{2}{*}{$6\times10^{-10}$} & \multirow{2}{*}{530} & 0 (standard candle) & $4\times10^{-3}$ & 0.04 & 0.5 & $6\times10^{-10}$\\
$\quad$\citet{ajello2014} & & & 1 (lognorm. $\sigma=1$) & $6\times10^{-3}$ & 0.11 & 1.1 & $10^{-10}$\\
\noalign{\vskip 2mm}
{\bf BL Lac objects} & \multirow{2}{*}{$2\times10^{-7}$} & \multirow{2}{*}{$1.2\times10^4$} & 0 & $1.9\times10^{-4}$ &  $6\times10^{-3}$ & 0.2  & $8\times10^{-9}$ \\
$\quad$\citet{ajello2014} & & & 1 & $3\times10^{-4}$ &  $0.03$ & 0.7 & $9\times10^{-10}$\\
\noalign{\vskip 2mm} 
{\bf Galaxy clusters} & \multirow{2}{*}{$3\times10^{-5}$} & \multirow{2}{*}{$1.9\times10^6$} & 0 & $1.1\times10^{-6}$ & $3\times 10^{-5}$ & $6\times10^{-3}$ & $2\times10^{-6}$\\
$\quad$\citet{zandanel2015} & & & 1 & $3\times10^{-6}$ & $5\times10^{-4}$ & $0.2$ & $1.4\times 10^{-7}$\\
\noalign{\vskip 2mm} 
{\bf Starburst galaxies} & \multirow{2}{*}{$3\times10^{-5}$} & \multirow{2}{*}{$1.8\times10^7$} & 0 & $1.3\times10^{-7}$ & $1.7\times10^{-6}$ & $3\times10^{-4}$ & $4\times10^{-5}$\\
$\quad$\citet{gruppioni2013} & & & 1 & $2\times10^{-7}$ & $3\times10^{-5}$ & $1.4\times10^{-2}$ & $2\times10^{-6}$\\
\hline                                  
\end{tabular}
\end{center}
}
\hfill{}
\end{table*}

The size of the bias depends on the number of sources in the population and on the cosmic source evolution and luminosity function. We quantify the bias for the different source classes listed in Table~\ref{tab:effective_densities} using the measured source rates and redshift distributions from the corresponding references. The probability distributions are calculated for each redshift distribution and the 90\% region and median of the probability distribution for one detected event are shown in the table. 

For each source class two different luminosity functions were adopted. The numbers in the upper line are for equally luminous sources, while a lognormal distribution with a width of one order of magnitude was assumed for the second line to show the impact of large luminosity fluctuations between individual sources. This variation in luminosity has the same impact on our results as using the observed distribution of gamma-ray luminosities for \textit{Fermi} LAT blazars which stretches over five orders of magnitudes (see Fig.~2 in \citealt{ajello2014}) and also reproduces fluctuations observed between individual gamma-ray burst by the \textit{Swift} Burst Alert Telescope \citep{wanderman2010}.
This simplistic treatment does not account for correlations between the source redshift and luminosity, which have been observed for example in galaxy clusters~\citep{gruppioni2013}. We find that the effect of the different redshift distributions and luminosity functions can be taken into account by using an effective density. The effective density is the density of a population of equally luminous sources without cosmic evolution which reproduces an equally large bias, i.e., the same median value. A similar calculation is performed in \citet{murase2016}, where the effective density quantifies the effect of the luminosity function, but not of the source evolution.

The last column of Table~\ref{tab:effective_densities} shows that the rather large luminosity fluctuations between individual sources reduce the effective source density by about one order of magnitude since a large fraction of the total neutrino flux is produced by a smaller number of sources. A detector that is not equally sensitive to all parts of the sky can be considered in the same way as the luminosity fluctuations since sources in the favored direction appear to be brighter. For example, if the detector is  only sensitive in half of the sky, the effective density is reduced by a factor of two and the bias is correspondingly smaller.

\begin{figure*}[tbp]
\begin{center}
\includegraphics[width=0.9\textwidth]{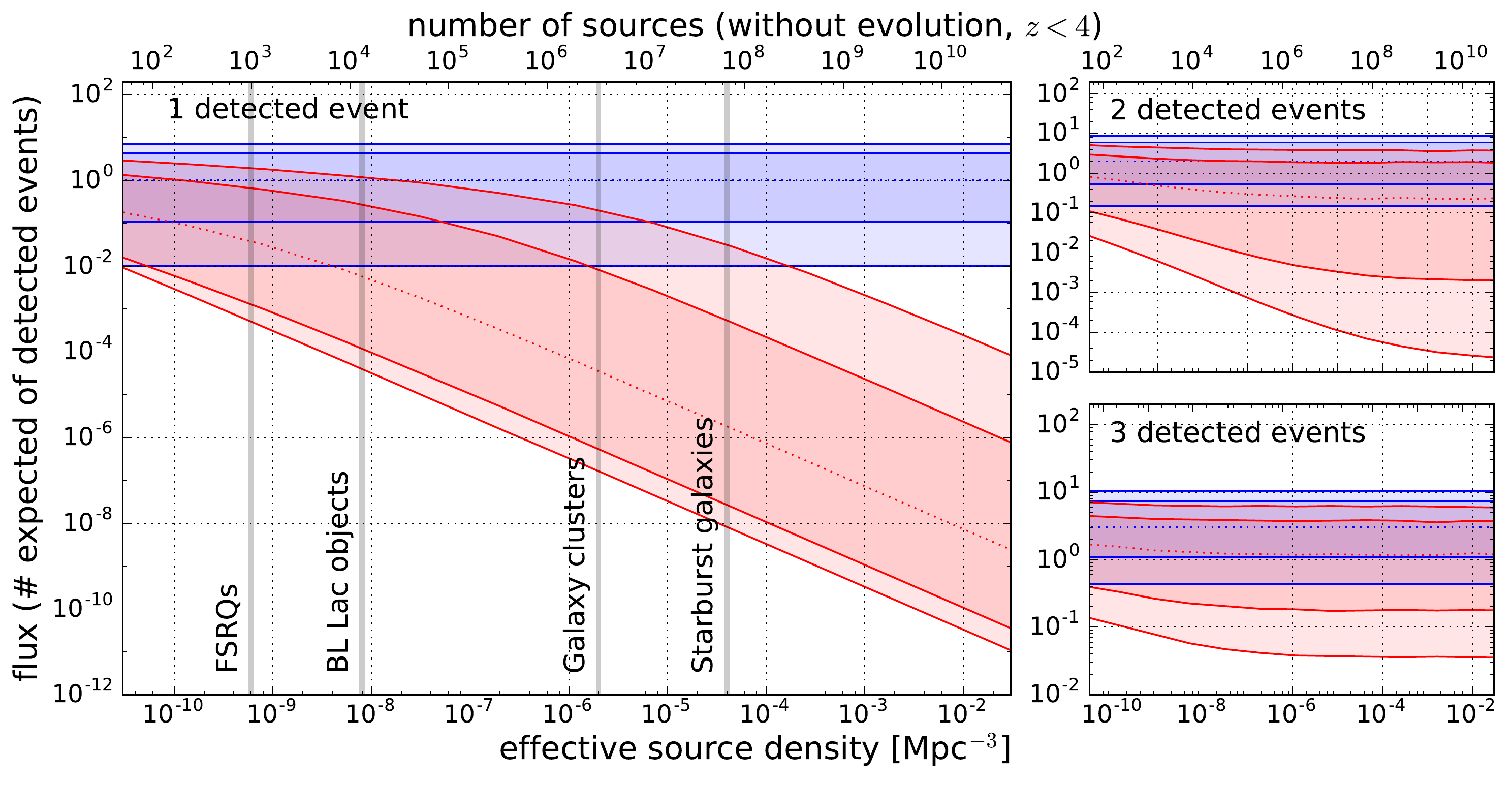}
\end{center}
\caption{Neutrino flux of a source detected with one (main panel), two, or three (smaller panels) neutrino events. The red bands show the real neutrino flux of the sources that produce the signal with 90\% and 99\% probability. For a single detected event, the source flux can be biased to lower values by several orders of magnitude compared to the flux inferred from the detected event (90\% and 99\% Poisson errors shown in blue). For this figure we assumed that the population in total produces 10 neutrino events (within a given time). While the selection bias is strong for sources detected with a single event, it quickly vanishes as soon as three or more events are detected from the same source.  The effective densities of several source classes are indicated (compare Table~\ref{tab:effective_densities}). \label{fig:comparison}}
\end{figure*}

Figure~\ref{fig:comparison} shows the size of the bias for different source densities assuming equally luminous sources and no cosmic evolution. The red bands show the expected neutrino flux of the sources where the dotted line indicates the median flux and the shaded area contains $90\%$ and $99\%$ of the probability distribution. 
The effective densities of the populations listed in Table~\ref{tab:effective_densities} are indicated in Fig.~\ref{fig:comparison};  the width of the inner red band approximately reproduces the size of the 90\% regions that were obtained for the different redshift distributions (see Table~\ref{tab:effective_densities}).


Especially for sources detected with just a single event, Fig.~\ref{fig:comparison} shows a strong discrepancy between the real flux (red bands) and the naive Poisson estimate (blue bands). The first row of Table~\ref{tab:effective_densities} shows that as few as 500 similar sources distributed throughout the universe reduce the median source luminosity by a factor of 20. However, this selection bias  quickly vanishes when three or more events are detected from the same source, as shown in the smaller panels of Fig.~\ref{fig:comparison}.





\section{Conclusions}

Our simulation shows that a single detected neutrino event is in general not sufficient to estimate the flux of a source. Depending on the number of undetected sources from the same population, the neutrino flux could be many orders of magnitude lower than the detected event seems to indicate. The bias also applies to transient (or variable) neutrino sources which could be supernovae, gamma-ray bursts, or the tidal disruption of a star.

When characterizing the properties of an individual potential neutrino source the consequences of the bias are easily overlooked. The bias is relevant whenever one or two events are associated with a source and are used to infer its neutrino flux, as was  done for example in \citet{krauss2014}, \citet{padovani2014}, \citet{petropoulou2015}, \citet{ptf12csypaper}, \citet{kadler2016}, \citet{padovani2016}, and \citet{gao2017}. In absence of a known source rate the neutrino flux estimates based on one or two events should therefore be considered upper limits (as  in, e.g., \citealt{txsmmpaper} or \citealt{padovani2018}).

The neutrino source candidate TXS\,$0506+056$ was initially found through the coincidence of a single high-energy neutrino event and a gamma-ray flare \citep{txsmmpaper}. When restricting the multiwavelength analysis to the properties of the observed gamma-ray flare, the neutrino luminosity thus cannot be estimated reliably and specific emission models should not be ruled out based on their low predicted neutrino flux (see, e.g.,~\citealt{gao2018, keivani2018}). The appropriate trial factor  here is not only the number of similar sources, but rather the rate density of similar blazar flares, and hence might be even larger than the effective density of BL Lac objects shown in Fig.~\ref{fig:comparison}.

There is evidence that additional neutrino events have been detected from TXS\,$0506+056$ \citep{txsneutrinopaper}. In this case the time-averaged source flux, as well as the neutrino flux during the 2014/2015 flare, can be measured accurately based on the large number of detected events. For a typical blazar detected with a single event the detection of further neutrinos seems unlikely due to the small fluxes shown in Fig.~\ref{fig:comparison}. However, one has to consider that for each detected EHE event tens to hundreds of lower energy events are expected depending on the spectral shape (compare the expected number of astrophysical events in \citealt{realtimepaper} and, e.g., \citealt{minutelongtransients}) which increases the probability of detecting further events correspondingly.

We summarize that the low number of astrophysical neutrino events detected so far leads to a strong selection effect caused by the Eddington bias. For a population as common as blazars, a source detected with a single event can be $1\,000$ times fainter than the neutrino detection seems to indicate (see Table~\ref{tab:effective_densities}). We emphasize that even a rather small population of similar sources can result in a strong bias, and the Eddington bias can only be neglected if the potential neutrino source is truly unique throughout the universe.

\begin{acknowledgements}
We thank Ludwig Rauch, Elisa Resconi, Andrea Palladino, Gabriel Espadas, and Paolo Padovani for the useful discussions. AF was supported by the Initiative and Networking Fund of the Helmholtz Association. 
\end{acknowledgements}

\bibliographystyle{aa}
\bibliography{eddingtonbias_references}

\end{document}